\documentstyle[prl,aps,epsfig,floats,multicol]{revtex}
\begin{document}
\preprint{
\vbox{\halign{&##\hfil         \cr
        & hep-ph/9911436      \cr
        & November 1999        \cr
        }}}

\title{Polarization of Prompt $\mbox{\boldmath$J/\psi$}$ at the Tevatron}

\author{Eric Braaten,$^1$ Bernd A. Kniehl,$^2$ and Jungil Lee$^2$}
\address{$^1$ Physics Department, Ohio State University, Columbus, OH 43210,
USA\\
$^2$ II. Institut f\"ur Theoretische Physik, Universit\"at Hamburg,
22761 Hamburg, Germany
}

\maketitle
\begin{abstract}
The polarization of prompt $J/\psi$ at the Fermilab Tevatron is calculated 
within the nonrelativistic QCD factorization framework.  
The contribution from radiative decays of P-wave charmonium states decreases,
but does not eliminate,
the transverse polarization at large transverse momentum.
The angular distribution parameter $\alpha$ for leptonic decays
of the $J/\psi$ is predicted to increase from near 0 at $p_T = 5$ GeV
to about 0.5 at $p_T = 20$ GeV.  
The prediction  is consistent with measurements by 
the CDF Collaboration at intermediate values of $p_T$, 
but disagrees by about  3 
standard deviations at the largest values of $p_T$ measured.

\smallskip
\smallskip
\smallskip
\noindent
PACS numbers: 13.85.t, 13.85.Ni, 14.40.Gx
\end{abstract}
\begin{multicols}{2}
The production of charmonium and bottomonium states 
in high-energy collisions probes
both the hard-scattering parton processes that create heavy
quark-antiquark ($Q\overline{Q}$) pairs 
and the hadronization process that transforms
them into color-singlet bound states.
One particularly sensitive probe is the polarization of the 
$J^{PC}=1^{--}$ quarkonium states.
The nonrelativistic QCD
(NRQCD) factorization approach to inclusive quarkonium production
\cite{B-B-L} makes the remarkable prediction that in hadron collisions 
these states should be transversely polarized at sufficiently large 
transverse momentum ($p_T$) \cite{Cho-Wise}.  Recent 
measurements at the Tevatron by the CDF Collaboration seem to be in dramatic
contradiction with this prediction~\cite{CDF-pol}.

As first pointed out by Cho and Wise \cite{Cho-Wise},
the prediction of transverse polarization 
for $1^{--}$ states at large $p_T$ 
follows from three simple features of the dynamics of massless 
partons and heavy quarks.  
First, the inclusive production of quarkonium 
(or any other hadron) at sufficiently large $p_T$
is dominated by {\it fragmentation}.
In $p \bar p$ collisions at the Tevatron,
the dominant contribution to the charmonium production rate at large $p_T$
comes from gluon fragmentation  \cite{B-Y}. 
The gluon is almost on shell and thus predominantly transversely polarized.
Second, a $Q \overline{Q}$ pair with small relative momentum 
created by the virtual gluon is, at leading order in $\alpha_s$, 
in a color-octet $^3S_1$ state  \cite{B-F} with the same 
transverse polarization as the gluon.
Third, the spin symmetry of nonrelativistic heavy quarks implies the
suppression of spin-flip transitions in the binding of the
$Q \overline{Q}$ pair into quarkonium.  
Thus, $1^{--}$ states should have a large transverse polarization
at sufficiently large $p_T$.  
A convenient measure of the polarization is the variable
$\alpha = (\sigma_T-2\sigma_L)/(\sigma_T+2\sigma_L)$, 
where $\sigma_T$ and $\sigma_L$  are the transverse 
and longitudinal components of the cross section, respectively.
Beneke and Rothstein studied the dominant fragmentation mechanisms for 
$\sigma_L$ \cite{Beneke-Rothstein}, and
concluded that, at sufficiently large $p_T$, 
$\alpha$ should be in the range 0.5 -- 0.8.

For charmonium production at the Tevatron, fragmentation does not yet
dominate for most of the $p_T$ range that is experimentally accessible.
In order to study the onset of the polarization effect,
it is necessary to take into account the {\it fusion} contributions
from parton processes $i j \to c \bar c + k$.
Quantitative calculations of the polarization variable $\alpha$
for {\it direct} $\psi'$ mesons ({\it i.e.} those that do not come from decays) 
at the Tevatron have been carried out by Beneke and
Kr\"amer \cite{Beneke-Kramer} and by Leibovich \cite{Leibovich}.
They predicted that $\alpha$ should be small for $p_T\alt 5$ GeV,
but then should rise dramatically
to  $0.77 \pm 0.08$ at $p_T =$ 20 GeV, according to Beneke and Kr\"amer,
and to $0.90 \pm 0.04$, according to Leibovich.
The CDF Collaboration has measured the polarization of direct
$\psi'$ \cite{CDF-pol},
but the error bars are too large to draw any definitive conclusions.  

The CDF Collaboration has also measured the 
polarization of {\it prompt} $J/\psi$ mesons \cite{CDF-pol} 
({\it i.e.} those that do not come from the decay of $B$ hadrons). 
The number of $J/\psi$ events
is larger than for $\psi'$ by a factor of about 100,
allowing $\alpha$ to be measured more precisely and in more $p_T$ bins.
They find that $\alpha$ has a positive value $0.32 \pm 0.10$
in the $p_T$ bin from 8 to 10 GeV. However, instead of
increasing at larger $p_T$, $\alpha$ decreases to 
$-0.29 \pm 0.23$ in the highest $p_T$ bin from 15 to 20 GeV.
Theoretical predictions of the polarization of prompt $J/\psi$
are complicated by the fact that the prompt signal includes
$J/\psi$ mesons that come from decays of the higher charmonium
states $\chi_{c1}$, $\chi_{c2}$, and $\psi'$. 
They account for about 15\%, 15\%, and 10\% of the prompt 
$J/\psi$ signal, respectively  \cite{CDF-old}.
The polarization of $J/\psi$ from $\psi'$ not via $\chi_{cJ}$ is 
straightforward to calculate, since the spin is unchanged by the transition.  
The polarization of $J/\psi$ from $\chi_{cJ}$ and of 
$J/\psi$ from $\psi'$ via $\chi_{cJ}$ is more complicated,
because the $\chi_{cJ}$ mesons are produced in various spin states 
and they decay into $J/\psi$ through radiative transitions. 

In this letter, we present a quantitative analysis of the polarization 
of prompt $J/\psi$ using the NRQCD factorization formalism.
We reanalyze the CDF data on the $p_T$ distributions for $J/\psi$,
$\chi_{c}$, and $\psi'$ to determine the relevant color-octet NRQCD matrix
elements (ME's).
The cross sections for the spin states of $J/\psi$, $\chi_{cJ}$,
and $\psi'$ are calculated using these ME's
and the appropriate parton cross sections.
The cross sections for the spin states of the $\chi_{cJ}$ 
required the calculation of new parton cross sections,
which will be published elsewhere \cite{chi-pol}.
The variable $\alpha$ is then obtained by combining these 
cross sections with the appropriate branching ratios into 
longitudinally polarized $J/\psi$ ($\psi_L$). 

The {\it NRQCD factorization formula} for the differential cross section for 
the inclusive production of a charmonium state $H$ of momentum $P$
and spin quantum number $\lambda$ has the schematic form
\begin{equation}
d \sigma^{H_\lambda(P)} \;=\;
d \sigma^{c \bar c_n(P)} \; 
	\langle O^{H_\lambda(P)}_n \rangle,
\label{sig-fact}
\end{equation}
where the summation index $n$ runs over
all the color and angular momentum states of the $c\bar c$ pair.
The $c \bar c$ cross sections $d \sigma^{c \bar c_n}$
can be calculated using perturbative QCD.
All dependence on the state $H$ is contained within the
nonperturbative ME's $\langle O^{H_\lambda(P)}_n \rangle$.
In general, they are Lorentz tensors that depend on the momentum $P$ 
and the polarization tensor of $H_\lambda$.
The Lorentz indices
are contracted with those of $d \sigma^{c \bar c_n}$
to give a scalar cross section.
The symmetries of NRQCD can be used to reduce 
the tensor ME's $\langle O^{H_\lambda(P)}_n \rangle$ 
to scalar ME's $\langle O^H_n \rangle$ that are
independent of $P$ and $\lambda$.
Thus one may calculate the  cross section for polarized quarkonium
once the relevant scalar ME's are known.
A nonperturbative analysis of NRQCD reveals
how the various ME's scale with the typical relative 
velocity of the heavy quarks.  
It also gives exact and approximate symmetry relations 
that can be used to simplify the ME's.
The most important ME's for the production of
$J/\psi=\psi(1S)$ or $\psi'=\psi(2S)$ 
can be reduced to one color-singlet parameter
$\langle O^{\psi(nS)}_1(^3S_1) \rangle$
and three color-octet parameters 
$\langle O^{\psi(nS)}_8(^3S_1) \rangle$,
$\langle O^{\psi(nS)}_8(^1S_0) \rangle$, and
$\langle O^{\psi(nS)}_8(^3P_0) \rangle$.
The most important ME's for $\chi_{cJ}$ production 
can be reduced to a color-singlet parameter
$\langle O^{\chi_{c0}}_1(^3P_0) \rangle$ 
and a single color-octet parameter 
$\langle O^{\chi_{c0}}_8(^3S_1) \rangle$.
The ME's enumerated above should be sufficient for a 
calculation of the polarization of prompt $J/\psi$.

In $p \bar p$ collisions,  the parton processes that dominate the
$c \bar c$ cross section depend on $p_T$.
If $p_T$ is of order $m_c$, those which dominate are {\it fusion} processes, 
whose contributions can be expressed as 
\begin{equation}
d \sigma^{H_\lambda(P)}_{\rm fu} =
f_{i/p} \otimes f_{j/\bar p}
\otimes
d \hat \sigma_{ij}^{c \bar c_n(P)} \;
        \langle O^{H_\lambda(P)}_n \rangle,
\label{sig-fusion}
\end{equation}
where $f_{i/p}(x,\mu)$ and $f_{j/\bar p}(x,\mu)$ are parton distribution
functions (PDF's) and a sum over the partons $i,j$ is implied.
The leading-order parton cross sections $d \hat \sigma$
are proportional to $\alpha_s^3(\mu)$.
These cross sections are given in Refs. \cite{Leibovich} 
and \cite{B-K-V} for all the relevant $c \bar c$ spin states
with the exception of color-singlet $^3P_J$ states,
which required a new calculation.
For $p_T\gg m_c$
the parton cross sections are dominated by 
{\it fragmentation} processes with the scaling behavior 
$d \hat \sigma/dp_T^2 \sim 1/p_T^4$.
These contributions can be expressed as 
\begin{equation}
d \sigma^{H_\lambda(P)}_{\rm fr}=
f_{i/p} \otimes f_{j/\bar p} \otimes
        d \hat \sigma_{ij}^{k(P/z)}
\otimes
D_k^{c \bar c_n} \; \langle O^{H_\lambda(P)}_n \rangle,
\label{sig-frag}
\end{equation}
where $D_k^{c \bar c_n}(z,\mu_{\rm fr})$ is a fragmentation function (FF).
We use a common renormalization and factorization scale $\mu$ for
$f_{i/p}$, $f_{j/\bar p}$, and $d \hat \sigma$,
but we allow $\mu_{\rm fr}$ to be different.
The momentum $k$ of the fragmenting parton is denoted
by $P/z$ in (\ref{sig-frag}).  
However, it is inconsistent to set $k^\mu=P^\mu/z$, 
since the parton is massless while $P^2 = 4 m_c^2$.
We choose $k^\mu$ so that $z$ is the fraction of the 
light-cone momentum of the parton $k$ that is carried by the 
$c \bar c$ pair in the parton CM frame.
The covariant expression is 
$k^\mu =
\left[(\Delta +K \cdot P) P^\mu - P^2 K^\mu \right]/(2 z \Delta)$,
where $K^\mu$ is the total momentum of the colliding partons $i$ and $j$,
and $\Delta = [(K \cdot P)^2 - K^2 P^2]^{1/2}$.

In order to predict the polarization of prompt $J/\psi$ at the Tevatron, 
we need values for the scalar ME's.
The color-singlet ME's $\langle O_1^{\psi(nS)}(^3S_1) \rangle$
and $\langle O^{\chi_{c0}}_1(^3P_0) \rangle$ can be determined 
phenomenologically from the decay rates for 
$\psi(nS)\to \ell^+\ell^-$ and $\chi_{c2}\to\gamma\gamma$ \cite{PDG}.
Using the vacuum saturation approximation and spin symmetry in the 
NRQCD factorization formulae and
including NLO QCD radiative corrections \cite{Barbieri},  
we obtain the values in Table \ref{TBL}.
The errors come from the experimental errors in the decay rates only.

The color-octet ME's are phenomenological
parameters that must be determined from production data. 
To predict the polarization at the Tevatron, 
it is preferable to use ME's extracted directly from 
Tevatron data in order to cancel theoretical errors
associated with soft gluon radiation.
There have been several 
previous extractions of the color-octet ME's 
\cite{Beneke-Kramer,Cho-Leibovich,C-G-M-P,Kniehl-Kramer}
from the CDF data on the $p_T$ distributions of 
$J/\psi$, $\chi_c$, and $\psi'$ \cite{CDF-old}.
We carry out an updated analysis largely following the strategy 
used in Ref.~\cite{Kniehl-Kramer}.
In the fusion cross section (\ref{sig-fusion}),
we include the parton processes $i j \to c \bar c + k$,
with $i,j=g,q,\bar q$ and $q = u,d,s$. 
In the fragmentation cross section (\ref{sig-frag}),
we include only the $g \to c \bar c_8(^3S_1)$ term, 
since this is the only fragmentation process for which
$D_k^{c \bar c_n}$ is of order $\alpha_s$.
The FF $D_g^{c\bar c_8(^3S_1)}$ is evolved in $\mu_{\rm fr}$ using the
standard homogeneous timelike evolution equation.
The effects of the violation of the phase-space constraint 
$\mu_{\rm fr}>4m_c^2/z$ are negligible at
the Tevatron due to the rapid fall-off of the $p_T$ distribution 
\cite{Kniehl-Zwirner}.

We consider two choices for the PDF's: MRST98LO as our default and CTEQ5L for
comparison \cite{PDF}.
We evaluate $\alpha_s$ from the one-loop formula 
using the value of $\Lambda_{\rm QCD}$ appropriate for the PDF set \cite{PDF}.
We set $\mu = (4 m_c^2 + p_T^2)^{1/2}$ and $m_c=1.5$ GeV.
The cross section for $\psi(nS)$ depends on the linear combination
$M_r = \langle O_8(^1S_0) \rangle + r \langle O_8(^3P_0) \rangle/m_c^2$,
where $r$ varies from about 3.6 at $p_T = 5.5$ GeV
to about 3.0 at $p_T = 18$ GeV, so we can only determine $M_r$
at some optimal value of $r$.
We determined $\langle O_8(^3S_1) \rangle$ and $M_r$ for $\psi(nS)$
by fitting the $p_T$ distributions from CDF
following the strategy in \cite{Kniehl-Kramer}.
We determined  $\langle O_8^{\chi_{c0}}(^3S_1) \rangle$
by fitting the $p_T$ distribution for $\chi_c$
together with the constraint from the preliminary CDF measurement of
$\sigma_{\chi_{c1}}/\sigma_{\chi_{c2}}$ \cite{CDF-old}.
Our values for the color-octet ME's are summarized in Table \ref{TBL}.
The error bars take into account the statistical errors only.
Our default $\psi'$ color-octet ME's agree within errors with those of
Ref.~\cite{Cho-Leibovich} used by Leibovich \cite{Leibovich} and with those
for 2 of the 3 PDF sets used by Beneke and Kr\"amer \cite{Beneke-Kramer}.
Our default $J/\psi$ color-octet ME's agree within errors with those of
Ref.~\cite{Cho-Leibovich}.
Our value for 
$\langle O_8^{J/\psi}(^3S_1) \rangle$ is about a factor of 3 smaller than
in Ref.~\cite{Beneke-Kramer}, 
while $M_r^{J/\psi}$ is about a factor of 2 larger.

We can calculate the cross sections for the polarized states $H_\lambda$
using the scalar ME's in Table \ref{TBL}.
The cross section (\ref{sig-fact})
can be reduced to an expression linear in the scalar ME's,
with coefficients that involve the
polarization tensor of $H_\lambda$.
In the channel $c \bar c_8(^3S_1)\to \psi_\lambda(nS)$, we interpolate
between the fusion cross section at low $p_T$ and the fragmentation
cross section at high $p_T$ using the prescription
\begin{equation}
d\sigma^{H_\lambda}
=
d\sigma^{H_\lambda}_{\rm fu}
\times
\left(
d\sigma^{H_\lambda}_{\rm fr}[\mu_{\rm fr}=\mu]
\big/
d\sigma^{H_\lambda}_{\rm fr}[\mu_{\rm fr}=2m_c]
\right).
\label{sig-interp}
\end{equation}
We proceed to summarize our calculation of the errors 
in $\sigma_L$ and $\sigma_T$.
The errors in the ME's in Table  \ref{TBL} are taken into account.
We take the central values of $\mu$  and $m_c$
to be $\mu_T = (4 m_c^2 + p_T^2)^{1/2}$ and $1.5$ GeV
and allow them to vary within the ranges
${1 \over 2}\mu_T$ -- $2\mu_T$ and  $1.45$ -- $1.55$ GeV, respectively.
We take MRST98LO as our default PDF, 
and we treat the difference between it and CTEQ5L as an error.
The cross section $\sigma_L$ for $\psi(nS)$ is sensitive to a different linear
combination of $\langle O_8(^1S_0) \rangle$ and
$\langle O_8(^3P_0) \rangle$ than appears in $M_r$.
We take this into account by expressing the cross section as a
function of $M_r$ and $x = \langle O_8(^1S_0) \rangle/M_r$,
taking the central value of $x$ to be ${1 \over 2}$,
and allowing $x$ to vary between 0 and 1.

We first consider the polarization of
direct $\psi'$, since it is not complicated by feeddown from 
higher charmonium states.
The polarization variable $\alpha$ measured by the 
CDF Collaboration \cite{CDF-pol}
describes the angular distribution of leptons from the decay 
of the $\psi'$ with respect to the $\psi'$ momentum in the 
hadron CM frame.    
The covariant expression for the polarization vector of $\psi'_L$ is
$(P^2 Q^\mu - P \cdot Q P^\mu)/(\sqrt{P^2} \Delta)$,
where $Q = p + \bar p$ is the total hadron momentum
and $\Delta =[(P \cdot Q)^2 - P^2 Q^2]^{1/2}$.
In Fig.~1(a), we compare our result for $\alpha$ 
as a function of $p_T$ with the CDF data \cite{CDF-pol}
and with previous predictions from Refs. \cite{Beneke-Kramer}
and \cite{Leibovich}.
We present our result in the form of an error band obtained by 
combining in quadrature all the errors described above.
The most important errors are those from
the $\psi^\prime$ ME's, the PDF's, and $x$.
The error bars in the CDF data are too large to draw any
definitive conclusions.
Our result for $\alpha$ is close to the prediction of Leibovich 
\cite{Leibovich}, and significantly larger than that
of Beneke and Kr\"amer \cite{Beneke-Kramer}.
Their calculations differ in the treatment
of terms of order $\alpha_s^2$ in the gluon FF \cite{Beneke-Rothstein}.
Beneke and Kr\"amer included these terms in $\sigma_L$ 
but neglected them in $\sigma_T$, while Leibovich neglected 
them in both $\sigma_L$ and $\sigma_T$.  We have adopted the strategy 
of Beneke and Kr\"amer, since these terms give a significant increase in 
$\sigma_L$ at large $p_T$ but have only a small effect on $\sigma_T$.
Although this tends to decrease $\alpha$, our smaller value of $M_r$
tends to increase $\alpha$, and the net result is close to the 
prediction of Leibovich.

We next consider the polarization variable $\alpha$ for prompt $J/\psi$.
The prompt cross section $\sigma_T + \sigma_L$
is the sum of the direct cross section for $J/\psi$ and 
the cross sections for $\chi_{cJ}$ and
$\psi'$ weighted by the branching fractions $B_{H \to J/\psi}$.
The prompt longitudinal cross section $\sigma_L$ 
is the sum of the direct cross section for $\psi_L$ and 
the cross sections for each of the spin states 
$\chi_{cJ(\lambda)}$ and $\psi'_\lambda$ 
weighted by $B_{H \to J/\psi}$
and by the probability $P_{H_\lambda \to \psi_L}$ 
for the polarized state to decay into $\psi_L$. 
The observed transitions of $\psi'$ to $J/\psi$ involve 
no spin flips, so that 
$P_{\psi'_\lambda \to \psi_L}$  is 1 for $\psi'_0$
and 0 for $\psi'_{\pm 1}$.
For the radiative decay of $\chi_{cJ(\lambda)}$ into $J/\psi$,
the probability $P_{\chi_{cJ(\lambda)} \to \psi_L}$ is
${1 \over 3}$ for $\chi_{c0}$, 
${1 \over 2}$ for $\chi_{c1(\pm 1)}$,
${2 \over 3}$ for $\chi_{c2(0)}$, ${1 \over 2}$ for $\chi_{c2(\pm 1)}$,
and 0 for the other spin states \cite{C-W-T}.
In Fig.~1(b), we compare our result for $\alpha$ 
as a function of $p_T$ with the CDF data \cite{CDF-pol}.
The shaded area indicates the 
error band obtained by adding the errors in quadrature.
The most important errors are those from the PDF's,
the $J/\psi$ ME's, and $x$. 
Our result for $\alpha$ is small around $p_T=5$ GeV, but it increases
with $p_T$ to a value around  0.5  at $p_T=20$ GeV.
Our result is in good agreement with the CDF measurement 
at intermediate values of $p_T$,
but it disagrees by about 3 standard deviations 
in the highest $p_T$ bin.
The three solid lines in Fig.~1(b) are the central curves of $\alpha$ for 
direct $J/\psi$, $J/\psi$ from $\chi_{cJ}$,
 and $J/\psi$ from $\psi'$. 
The $\alpha$ for direct $J/\psi$ is smaller than that
for direct $\psi'$, because
$\langle O_8(^3S_1) \rangle$ is comparable for $J/\psi$ and $\psi'$,
while $M_r$ is significantly larger for $J/\psi$.
In the moderate-$p_T$ region,
the contributions from $\psi'$ and from $\chi_c$ add to give an increase 
in the transverse polarization of prompt $J/\psi$ 
compared to direct $J/\psi$.  In the high-$p_T$ region,
the contributions from $\psi'$ and $\chi_c$ tend to cancel.
The prediction of Beneke and Kr\"amer for $\alpha$ for direct $J/\psi$ 
is identical to their prediction for direct $\psi'$ in Fig.~1(a). 
At $p_T = 20$ GeV, it is 
significantly larger than our prediction for direct $J/\psi$.
The difference comes from our smaller value for 
$\langle O_8^{J/\psi}(^3S_1) \rangle$ 
and our larger value for $M_r^{J/\psi}$.
Beneke and Kr\"amer's prediction for $\alpha$ for 
$J/\psi$ from $\psi'$ would be significantly lower than our result in 
Fig.~1(b), but it would have a small effect on the value of $\alpha$ 
for prompt $J/\psi$.  The discrepancies between their predictions 
and ours could be eliminated 
by more accurate data on the $J/\psi$ and $\psi'$ cross sections, 
which would decrease some of the ambiguities in the analysis.

The CDF measurement of the polarization of prompt $J/\psi$
presents a serious challenge to the NRQCD factorization formalism
for inclusive quarkonium production.
There are many effects that could change our quantitative prediction
for $\alpha$, 
such as  next-to-leading order radiative corrections, but
the qualitative prediction that $\alpha$ should increase
at large $p_T$ seems inescapable.
In Run II of the Tevatron, the data sample for $J/\psi$ should be 
more than one order of magnitude larger than in Run I, 
allowing the polarization to be measured with higher 
precision and out to larger values of $p_T$.
If the result continues to disagree with the predictions 
of the NRQCD factorization approach, it would indicate a serious flaw in our 
understanding of inclusive charmonium production.
The predictions of low-order perturbative QCD for the 
spin-dependence of $c \bar c$ cross sections could be wrong, 
or the use of NRQCD to understand the systematics
of the formation of charmonium from the $c \bar c$ pair could be flawed,
or $m_c$ could simply be too small to apply
the factorization approach to the charmonium system.

J.L. thanks Sungwon Lee for  the graphics.
This work was supported in part 
by DOE  Grant No. DE-FG02-91-ER40690,
by DFG  Grant No. KN 365/1-1,
by BMBF Grant No. 05 HT9GUA 3,
through TMR Network No. ERBFMRX-CT98-0194,
and by Humboldt Foundation.

\end{multicols}
\begin{table}
\begin{tabular}{ccccccccccc}
PDF      &
         $\langle O^{J/\psi}_1(^3S_1)\rangle$&
         $\langle O^{J/\psi}_8(^3S_1)\rangle$&
         $M_{3.4}^{J/\psi}$& 
         $\langle O^{\psi^\prime}_1(^3S_1)\rangle$&
         $\langle O^{\psi^\prime}_8(^3S_1)\rangle$&
         $M_{3.5}^{\psi^\prime}$& 
         $\langle O^{\chi_{c0}}_1(^3P_0)\rangle$&
         $\langle O^{\chi_{c0}}_8(^3S_1)\rangle$
\\
MRST98LO&
$1.3\pm 0.1$&
$4.4\pm 0.7$&
$8.7\pm 0.9$&
$6.5\pm 0.6$&
$4.2\pm 1.0$&
$1.3\pm 0.5$&
$8.9\pm 1.3$&
$2.3\pm 0.3$
\\
CTEQ5L&
$1.4\pm 0.1$&
$3.9\pm 0.7$&
$6.6\pm 0.7$&
$6.7\pm 0.7$&
$3.7\pm0.9$&
$0.78\pm 0.36$&
$9.1\pm 1.3$&
$1.9\pm 0.2$
\\
unit&
         GeV$^3$&
         $10^{-3}$GeV$^3$&
         $10^{-2}$GeV$^3$&
         $10^{-1}$GeV$^3$&
         $10^{-3}$GeV$^3$&
         $10^{-2}$GeV$^3$&
         $10^{-2}$GeV$^5$&
         $10^{-3}$GeV$^3$
\end{tabular}
\caption{
NRQCD matrix elements.
The error bars take into account the statistical errors only.
}
\label{TBL}
\end{table}
\begin{figure}
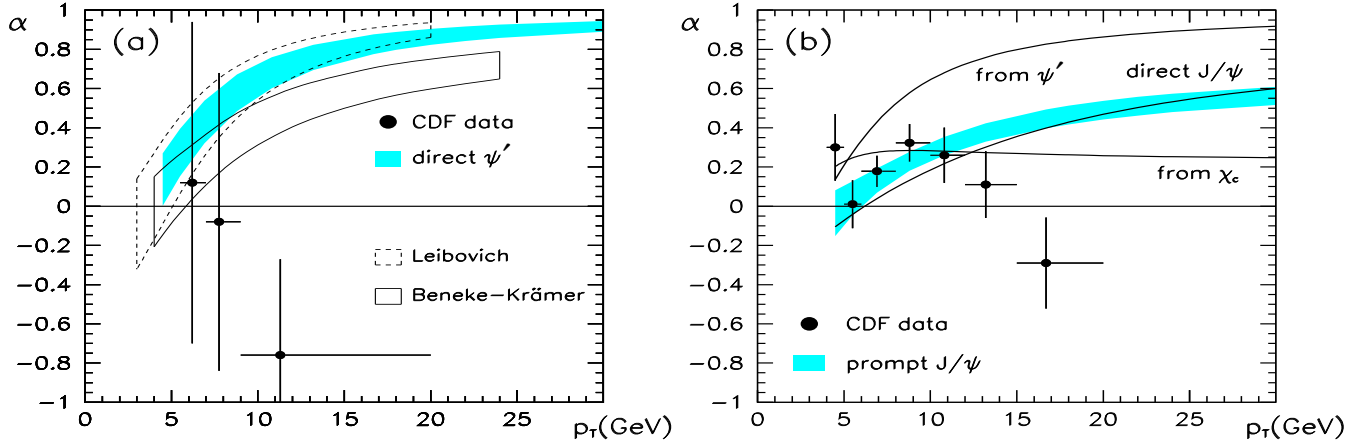

\begin{center}
\begin{tabular}{cc}
\epsfig{file=a_dr2bw.epsi,width=8.8cm,height=5.8cm}&
\epsfig{file=a_psibw.epsi,width=8.8cm,height=5.8cm}
\end{tabular}
\end{center}
\caption{
Polarization variable $\alpha$ vs.\ $p_T$
for (a) direct $\psi'$ and (b) prompt $J/\psi$
compared to CDF data.
}
\end{figure}
\end{document}